\documentclass[a4paper,fleqn,usenatbib]{mnras}

\usepackage{newtxtext,newtxmath}

\usepackage[T1]{fontenc}
\usepackage{ae,aecompl}

\usepackage{graphicx}	
\usepackage{amssymb}	
\usepackage{graphics}

\def\apj{ApJ}
\def\apjs{ApJS}
\def\mnras{MNRAS}

\def\apjl{ApJ}
\def\aap{A\&A}
\def\aj{AJ}

\def\apss{Ap\&SS}

\def\asec{\ifmmode ^{\prime\prime}\else$^{\prime\prime}$\fi}

\def\it{\sl}
\def\degs{\ifmmode ^{\circ}\else$^{\circ}$\fi}
\def\amin{\ifmmode ^{\prime}\else$^{\prime}$\fi}
\def\asec{\ifmmode ^{\prime\prime}\else$^{\prime\prime}$\fi}
\def\farcs{\hbox{$.\!\!^{\prime\prime}$}}  

\def\degs{\ifmmode ^{\circ}\else$^{\circ}$\fi}
\def\amin{\ifmmode ^{\prime}\else$^{\prime}$\fi}

\def\eqalign#1{\null\,\vcenter{\openup1\jot \m@th
   \ialign{\strut\hfil$\displaystyle{##}$&$\displaystyle{{}##}$\hfil
   \crcr#1\crcr}}\,}

\title[X-ray counterpart candidates for six  new $\gamma$-ray  pulsars]{X-ray counterpart candidates for six new $\gamma$-ray pulsars}

\author[D. A. Zyuzin et al.]{
Dmitry A. Zyuzin,$^{1}$\thanks{E-mail: da.zyuzin@gmail.com (DAZ)}
Anna V. Karpova,$^{1}$
and Yura A. Shibanov$^{1}$
\\
$^{1}$Ioffe Institute, Politekhnicheskaya ul., 26, St. Petersburg, 194021,  Russia
}

\date{Accepted XXX. Received YYY; in original form ZZZ}

\pubyear{2017}

\begin{document}
\label{firstpage}
\pagerange{\pageref{firstpage}--\pageref{lastpage}}
\maketitle

\begin{abstract}
Using archival X-ray data we have  found point-like X-ray counterpart candidates positionally 
coincident  with  six $\gamma$-ray pulsars discovered recently 
in the \textit{Fermi} Gamma-ray Space Telescope data  by the Einstein@Home project.
The  candidates for PSRs J0002$+$6216, J0554$+$3107, J1844$-$0346 and 
J1105$-$6037 are detected  with \textit{Swift}, and those for PSRs J0359$+$5414 and J2017$+$3625 are detected  with  \textit{Chandra}.  
Despite a low count statistics for some candidates, assuming plausible constraints on the absorbing column density towards the pulsars,  
we show that X-ray spectral properties for all of them  are consistent with those observed for other pulsars. J0359$+$5414 
is the most reliably identified object.  
We detect a nebula around it,
whose   spectrum and extent suggest that this is  a pulsar wind nebula powered by the  pulsar. 
Associations of  J0002$+$6216 and J1844$-$0346 with  supernova remnants CTB 1 and G28.6$-$0.1 are proposed.
\end{abstract}



\section{Introduction}
The Large Area Telescope (LAT) aboard the \textit{Fermi} Gamma-ray Space Telescope has 
discovered over 200 new 
young and millisecond  $\gamma$-ray 
pulsars\footnote{https://confluence.slac.stanford.edu/display/GLAMCOG/
Public+List+of+LAT-Detected+Gamma-Ray+Pulsars}. 
The inferred parameters for the 
young pulsars characterise most of them 
as relatively energetic    
rotation powered nearby neutron stars (NSs). With spin-down powers $\dot{E} \gtrsim 10^{34}$ erg s$^{-1}$  
and characteristic ages $\tau_c \lesssim 1$ Myr   they are likely 
located in less than a few kpc from the Sun \citep{abdo2013}. 
Young  NSs, in particular nearby ones, are rare  \citep[e.g.,][]{noutsos2013}.
As such, these discoveries 
contribute significantly to more complete understanding of the pulsar population, 
NS physics, and birthrates \citep{2011watters}.
About a third of the LAT pulsars was detected during `blind' searches for 
periodicity in sparse LAT $\gamma$-ray photons.  
Many of them remain undetected in the radio precluding  independent 
distance estimates using the dispersion measure. 
In this case, follow-up observations at other wavelengths, 
especially in X-rays, were proven to be very productive \citep[e.g.,][]{2013marelli, marelli2015}. 

More distant and/or less energetic pulsars are weaker in $\gamma$-rays and 
long integration times are required for 
a detectable signal-to-noise ratio. 
The photon 
sparseness
for them 
results 
in  a large computational cost of the blind search. 
The problem is solved with the aid of 
the volunteer-based computing
power of the Einstein@Home project\footnote{http://einstein.phys.uwm.edu} \citep{2013allen}. 
Within this project, 19 new $\gamma$-ray pulsars have been recently discovered in 
the blind search of pulsations in  about hundred of unidentified 
point-like LAT sources whose $\gamma$-ray spectra are similar 
to those of pulsars  \citep{Pletsch2013, clark2015, clark2016, Clark2017}. 
All of them are isolated  pulsars with $\tau_c$ between 3 kyr and 2 Myr and 
$\dot E$ between $10^{34}$ and $4\times 10^{36}$ erg s$^{-1}$.
No radio pulsations have been reported so far for any of these new pulsars.   
Therefore, X-ray identification of these objects is crucial for their further study. 
\begin{table*}
\begin{center}
\caption{X-ray counterpart candidates for six \textit{Fermi} pulsars  from the Einstein@Home sample. 
$\gamma$-ray coordinates are  from \citet{Pletsch2013, Clark2017} and 
X-ray coordinates are  from the 1st \textit{Swift} X-ray Point Source Catalogue \citep[1SXPS;][]{SwiftCatalogue} 
and the \textit{Chandra} Source Catalogue  \citep{ChandraCatalogue}.  
Numbers in parentheses are 1$\sigma$ uncertainties for the $\gamma$-ray coordinates and 90 per cent confidence position errors for the X-ray counterpart 
candidates related to 
the last significant digits quoted. The last column shows significances of the X-ray candidate detections as the signal to noise ratio $S/N$.  
}
\footnotesize
\begin{tabular}{ccccccc}
\hline
\textit{Fermi} Pulsar & X-ray Counterpart Candidate  &  RA$_{\gamma}$ & Dec$_{\gamma}$ & RA$_{\rm X}$ & Dec$_{\rm X}$ & $S/N$  \\ 
\hline
J0002$+$6216 & 1SXPS J000257.6+621609        & 00:02:58.17(2)    & +62:16:09.4(1)    & 00:02:57.69(70)   & +62:16:09.2(4.9) & 1.8\\
J0359$+$5414 & CXOGSG J035926.0+541455   & 03:59:26.01(2)    & +54:14:55.7(3)    & 03:59:26.09(11)   & +54:14:55.8(1.0) & 23.8 \\
J0554$+$3107 & 1SXPS J055404.8+310741        & 05:54:05.01(3)    & +31:07:41(4)       & 05:54:04.83(91)   & +31:07:41.7(6.2) & 2.7  \\
J1105$-$6037  & 1SXPS J110500.3-603713         & 11:05:00.48(4)    & -60:37:16.3(3)      & 11:05:00.37(84)   & -60:37:13.1(6.4) & 2.0 \\
J1844$-$0346  & 1SXPS J184432.9-034626         & 18:44:32.89(2)    & -03:46:30.6(9)     & 18:44:32.9(2)     & -03:46:26.6(2.7) & 5.9 \\
J2017$+$3625  & CXOGSG J201755.8+362507  & 20:17:55.84(1)    & 36:25:07.9(2)       & 20:17:55.81(8)   & 36:25:07.8(1.0) & 5.5 \\

\hline
\end{tabular}
 \label{t:psr-coo}
 \end{center}
\end{table*}
\begin{table*}
\begin{center}
\caption{Parameters of the pulsars listed in Table~\ref{t:psr-coo}, which are derived 
from the \textit{Fermi}-LAT timing solutions \citet{Pletsch2013, Clark2017}: 
a period $P$, a period derivative $\dot{P}$, a characteristic age $\tau_c=P/2\dot{P}$, 
a spin-down luminosity $\dot{E}$, a dipole magnetic field $B$.  
The \textit{Fermi}-LAT flux density $G_{100}$ in the  100 MeV--100 GeV range  
is taken from 3FGL catalogue \citep{3fgl}. 
$d_{\gamma}$ is a pseudo--distance (see text for details).} 
\begin{tabular}{cccccccc}
\hline
\textit{Fermi} Pulsar & $P$  &  $\dot{P}$ & $\tau_{\rm c}$  &  $\dot{E}$     & $B$ & $G_{100}$         & $d_{\gamma}$ \\
           & (ms) &  (10$^{-15}$ s s$^{-1}$)                               & (kyr)                  &  ($10^{33}$ erg s$^{-1}$)  & (10$^{12}$ G)   & (10$^{-11}$ erg cm$^{-2}$s$^{-1}$) & (kpc)  \\
\hline
J0002$+$6216 & 115.4 & 5.97   & 306 & 153   & 0.8 & 1.83 & 2.3   \\
J0359$+$5414 & 79.4   & 16.7   & 75   & 1318 & 1.2 & 2.45 & 3.45 \\
J0554+3107     & 465.0    & 142.6    & 52   & 56     &  8.2   & 1.73 & 1.9   \\
J1105$-$6037  & 194.94  & 21.8 & 141 & 116 & 2.1 & 3.7 & 1.53 \\
J1844$-$0346  & 112.85 & 154.7  & 12     &4249 & 4.2    & 2.83       & 4.3 \\
J2017$+$3625 & 166.75 & 1.36   &  1943 &12     &0.5 & 6.5  & 0.656 \\
\hline
\end{tabular}
\label{t:pulsars}
\end{center}
\end{table*}

Using archival \textit{Swift} and \textit{Suzaku} data we have found 
a plausible  X-ray counterpart and  a possible pulsar wind nebula (PWN)    
for PSR J1932$+$1916, which is one of the youngest ($\tau_c\approx 35.4$ kyr) pulsars 
from the Einstein@Home sample \citep{2017karpova}. 
The X-ray data allowed us to constrain the distance to the pulsar and 
to suggest its possible association with the supernova remnant (SNR) G54.4--0.3.

Here we report the  results of search for X-ray counterparts for 
the other eighteen Einstein@Home $\gamma$-ray pulsars using X-ray data archives and point source catalogues. 
We have found six counterpart candidates  by  position coincidence.   
We focus on their X-ray spectral properties 
and discuss if they are consistent with  those of other \textit{Fermi} pulsars 
which have been firmly identified in X-rays.  
Possible  associations with nearby SNRs are proposed for two pulsars  which allow for independent distance estimates. 
In Sect.~\ref{search} we present the search results and describe the general techniques and tools used for the X-ray analysis of the counterpart candidates.
The X-ray properties of each candidate and the possible SNR associations are thoroughly considered in sub-sections of Sect.~\ref{prop}.   
Background optical objects within the position error ellipses of the candidates 
are considered in Sect.~\ref{oc}. 
In Sect.~\ref{dc},  we discuss the results.
  
 \section{X-ray counterpart search results}
 \label{search}

Using X-ray catalogues\footnote{https://heasarc.gsfc.nasa.gov/w3browse/all/xray.html;\\
http://www.swift.ac.uk/1SXPS/}, we have found point-like X-ray counterpart candidates for 
six out of eighteen Einstein@Home  $\gamma$-ray pulsars, which  coincided  by position with 
the \textit{Fermi} pulsars at 90\% confidence level. 
Four candidates have been detected  with \textit{Swift} X-Ray Telescope (XRT) and the other two have been detected 
with  \textit{Chandra} Advanced CCD Imaging Spectrometer (ACIS).    
The \textit{Fermi} pulsar and X-ray counterpart candidate catalogue names 
and respective $\gamma$- and X-ray coordinates with their uncertainties are presented in Table~\ref{t:psr-coo}. 
The parameters of the pulsars derived from the $\gamma$-ray data are shown in Table~\ref{t:pulsars}. 
For radio-quiet pulsars the distances are estimated
using the empirical `pseudo'-distance relation for $\gamma$-ray pulsars \citep{SazParkinson2010}: 
$d_\gamma = 1.6\times(\dot{E}{\rm [10^{34}\ erg~s^{-1}]})^{0.25}\times
(G_{100}{\rm [10^{-11}\ erg~cm^{-2}~s^{-1}]})^{-0.5}$ kpc,
where $G_{100}$ is the pulsar flux density in the   100 MeV--100 GeV range 
measured with the \textit{Fermi}-LAT.

In most cases, the regions containing the pulsars were observed in X-rays several times. 
For  \textit{Swift} candidates, the number of observations varies 
 from 2, for J0554$+$3107,  to 64, for J1844$-$0346,  and the total exposure times varies from 9.2, for J0002$+$6216,  to 100 ks, for J1844$-$0346, respectively. 
For  \textit{Chandra} candidates,  they varied from 1 to 9 and from 10 ks to  460 ks, for J2017$+$3625\footnote{Observation ID  14699} and  
J0359$+$5414\footnote{Observation IDs 4657, 14688, 14689, 14690, 15548, 15549, 15550, 15585, 15586}, respectively. 
In Table~\ref{t:psr-coo}, the coordinates of the \textit{Swift} candidates are taken from the 1st \textit{Swift} X-ray Point Source Catalogue  
\citep{SwiftCatalogue}. They were measured on the stacked XRT images. 
For \textit{Chandra}  objects,  the archival ACIS data were reprocessed using the CIAO v.4.9 tool {\sc chandra\_repro}.  
For  multiple observations, the data were 
merged using  the {\sc merge\_obs} task. The candidate coordinates were measured on the resulting images using the {\sc wavdetect} tool. 
They are consistent with those provided by the \textit{Chandra} Source Catalogue  \citep{ChandraCatalogue}.  
\begin{figure*} 
 \setlength{\unitlength}{1mm}  
  \begin{center}
   \begin{picture}(145,215)(0,0)
    \put(0,140){\includegraphics[scale=0.25, angle=0,bb=58 150 871 953,clip]{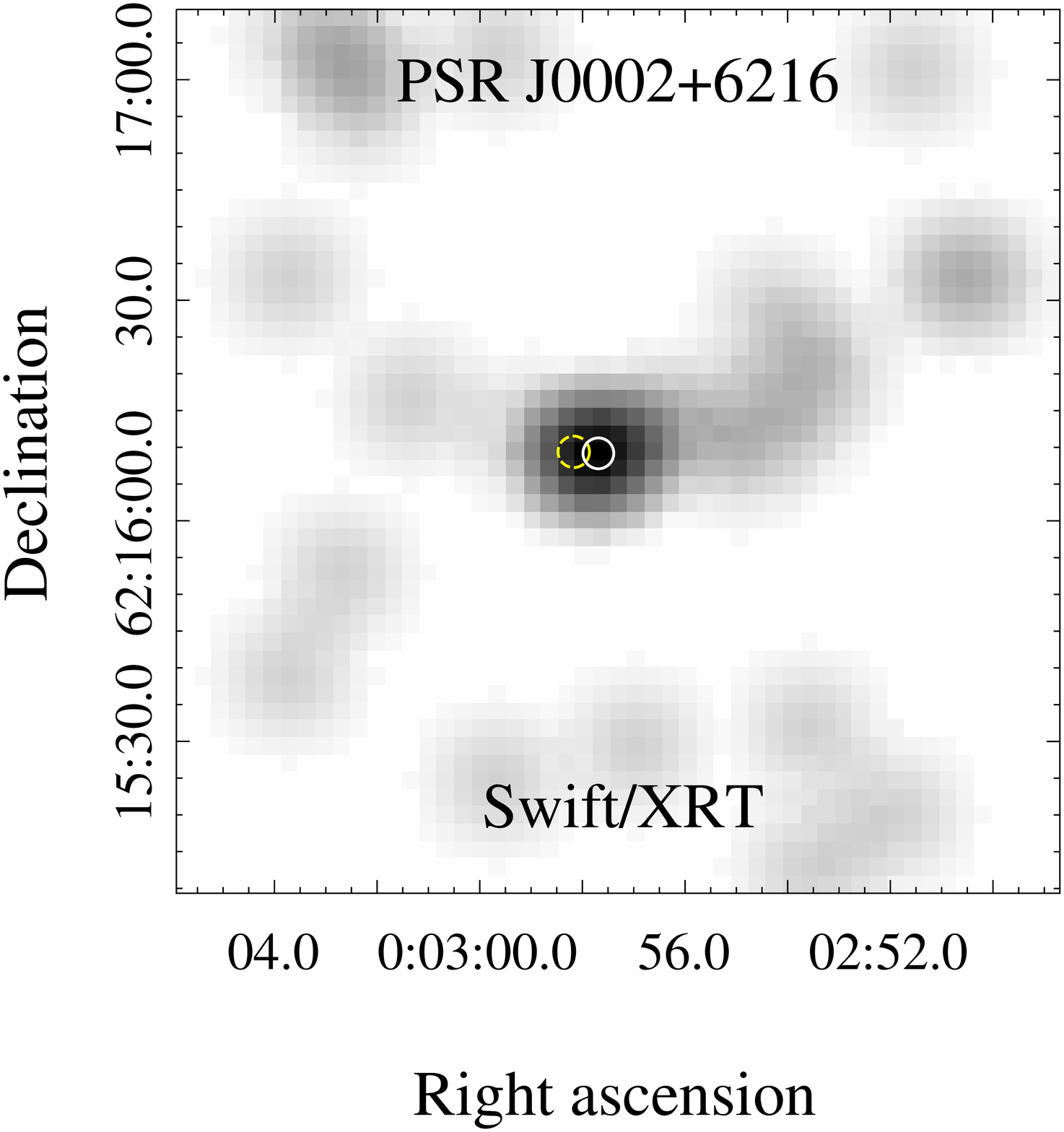}}
    \put(80,140){\includegraphics[scale=0.245, angle=0,bb=167 166 896 903,clip]{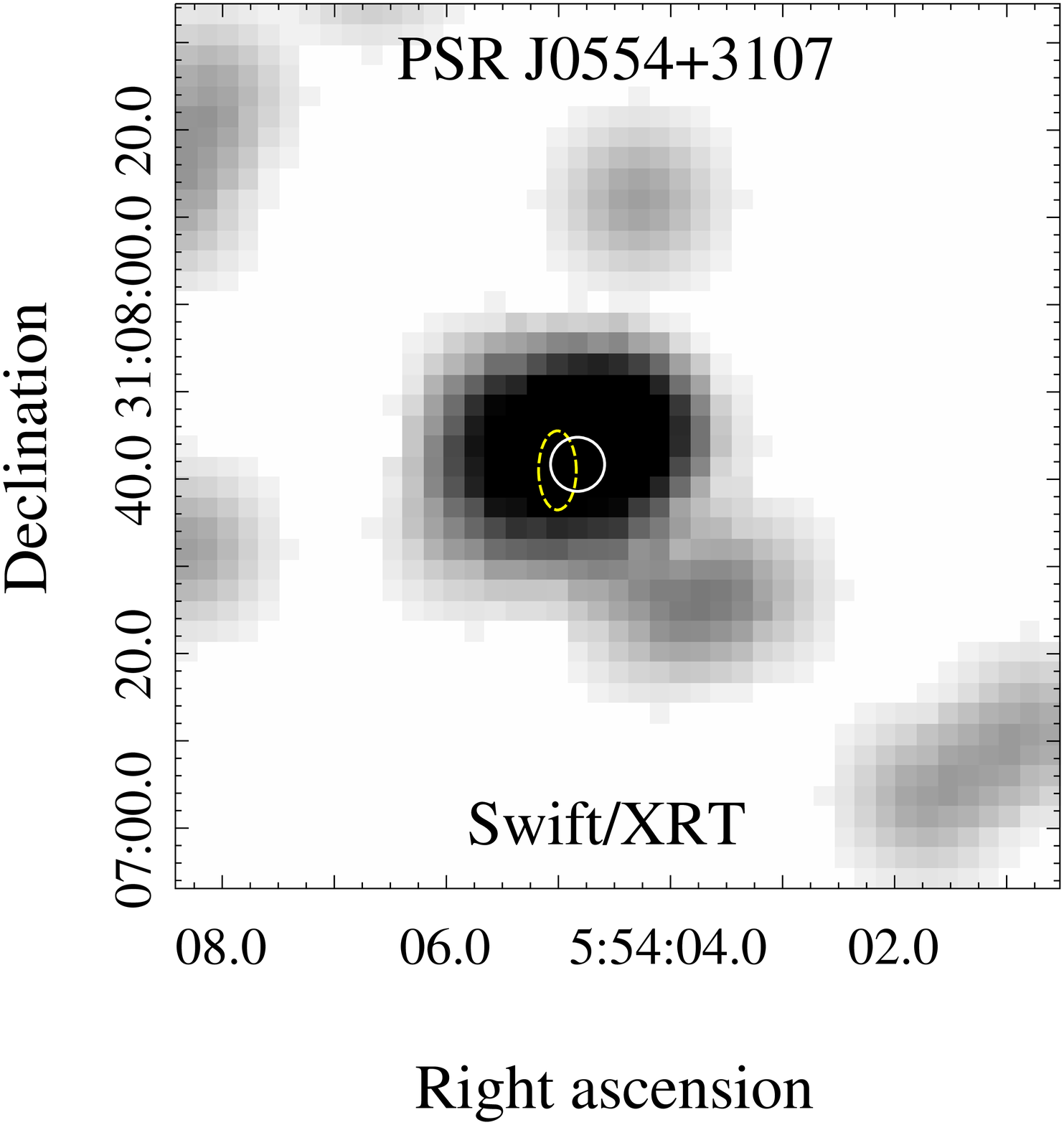}}
    \put(-1,72){\includegraphics[scale=0.25, angle=0,bb=83 170  892 897,clip]{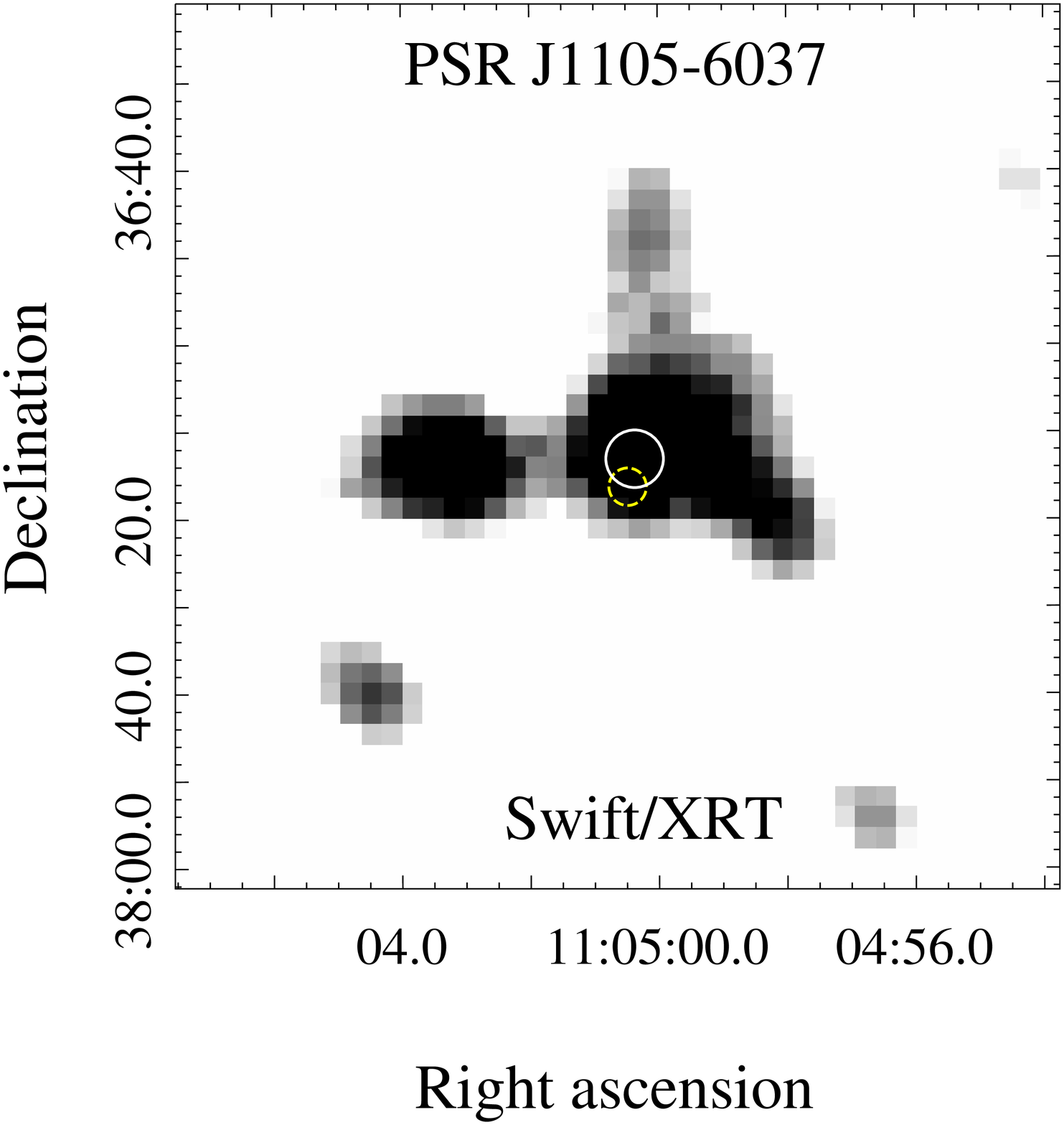}}
    \put(80,72){\includegraphics[scale=0.25, angle=0,bb=146 164 929 930,clip]{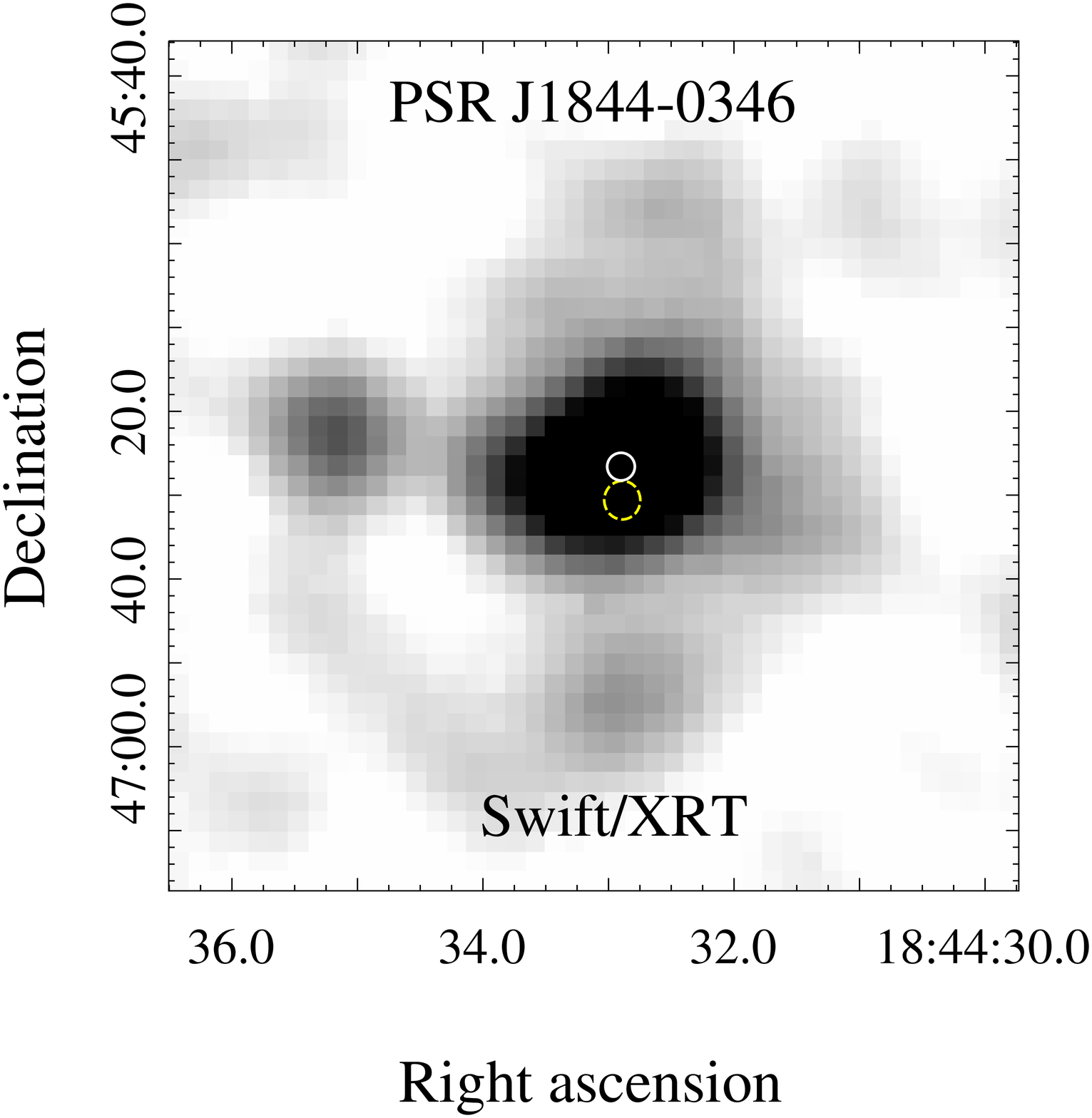}}
    \put(0,-2){\includegraphics[scale=0.225,bb=28 43 930 933, clip]{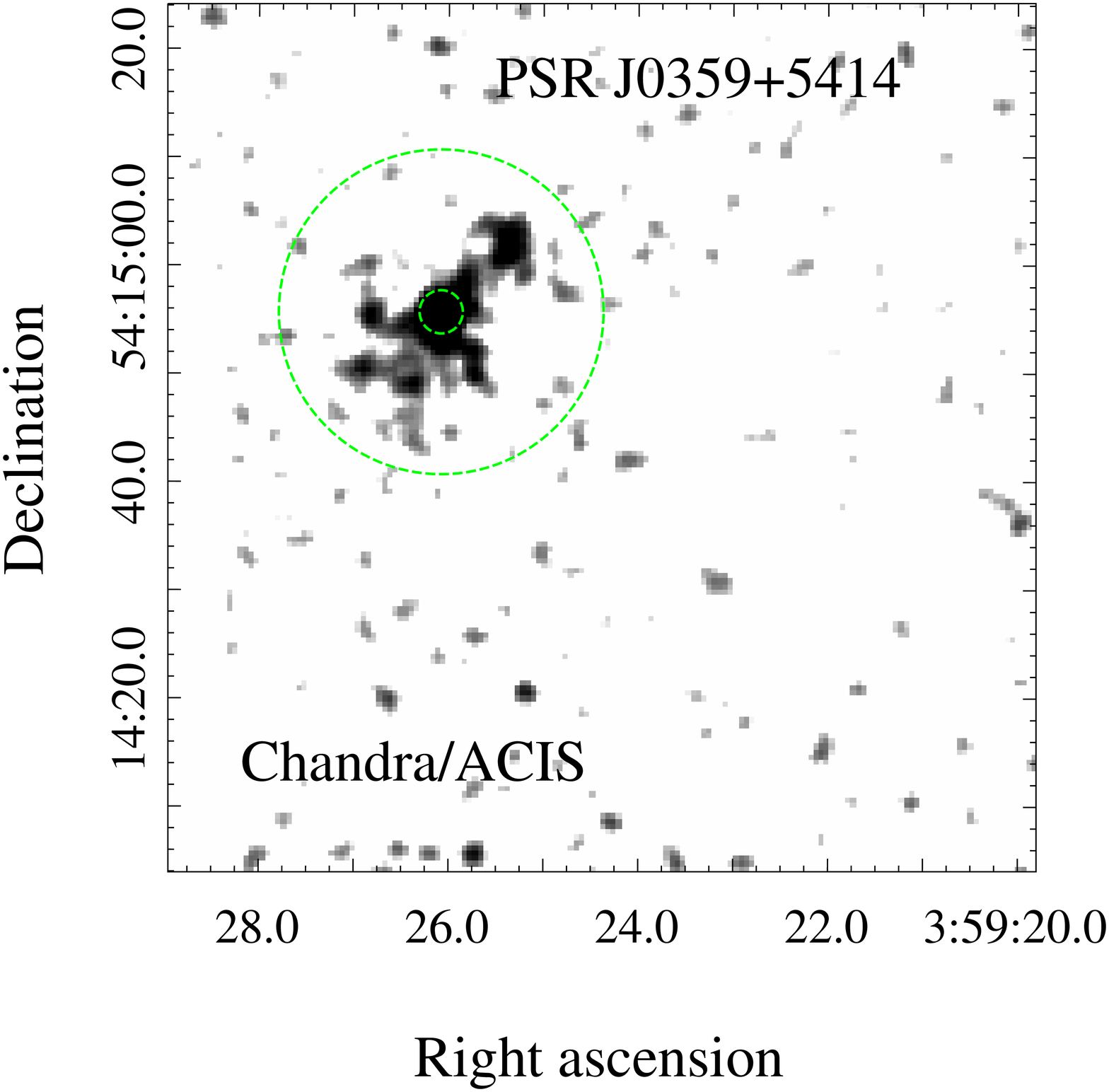}}
    \put (44,13) {\includegraphics[scale=0.17, clip]{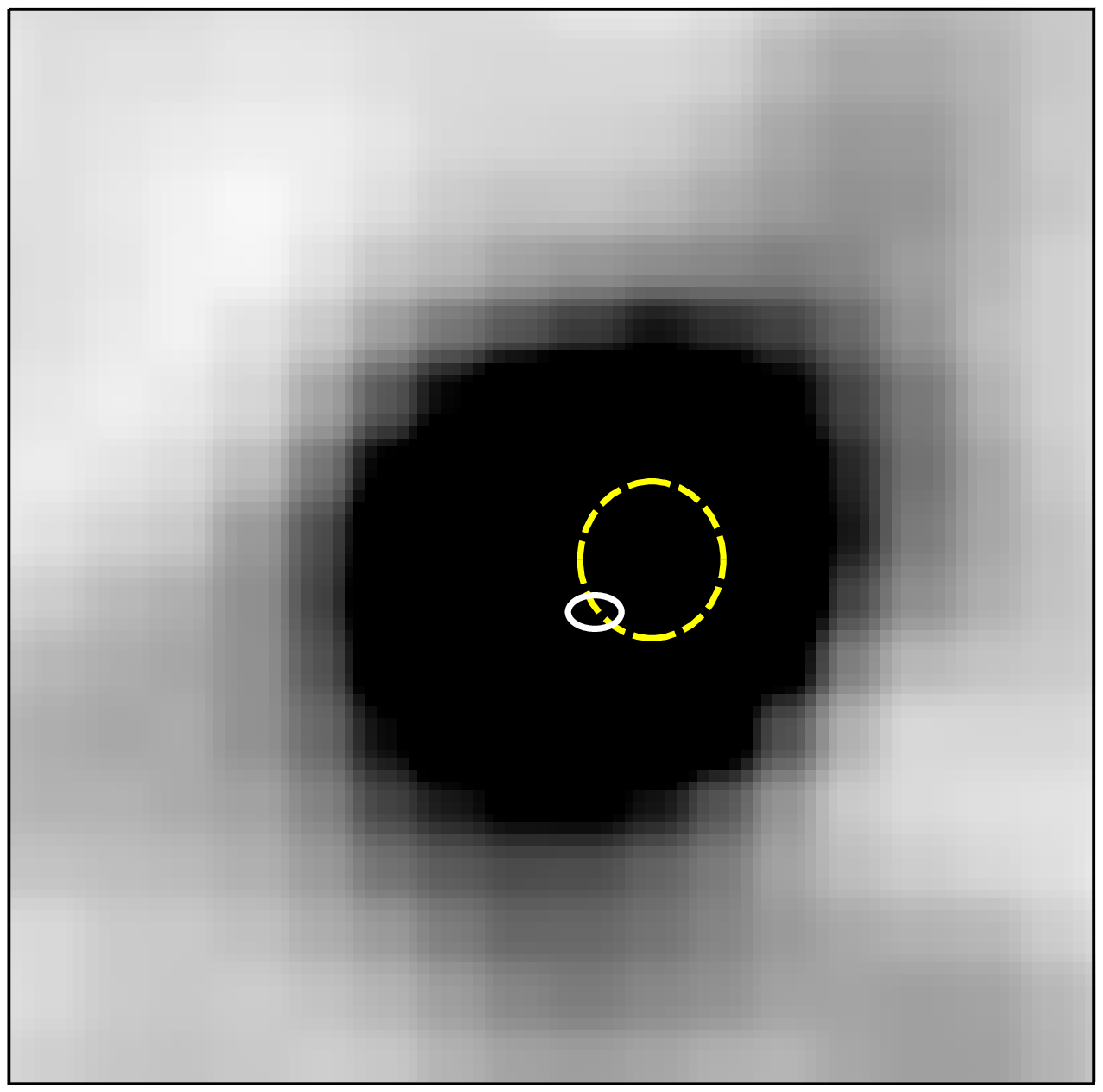}} 
    \put (82,-1) {\includegraphics[scale=0.23, bb=168 11 878 853, clip]{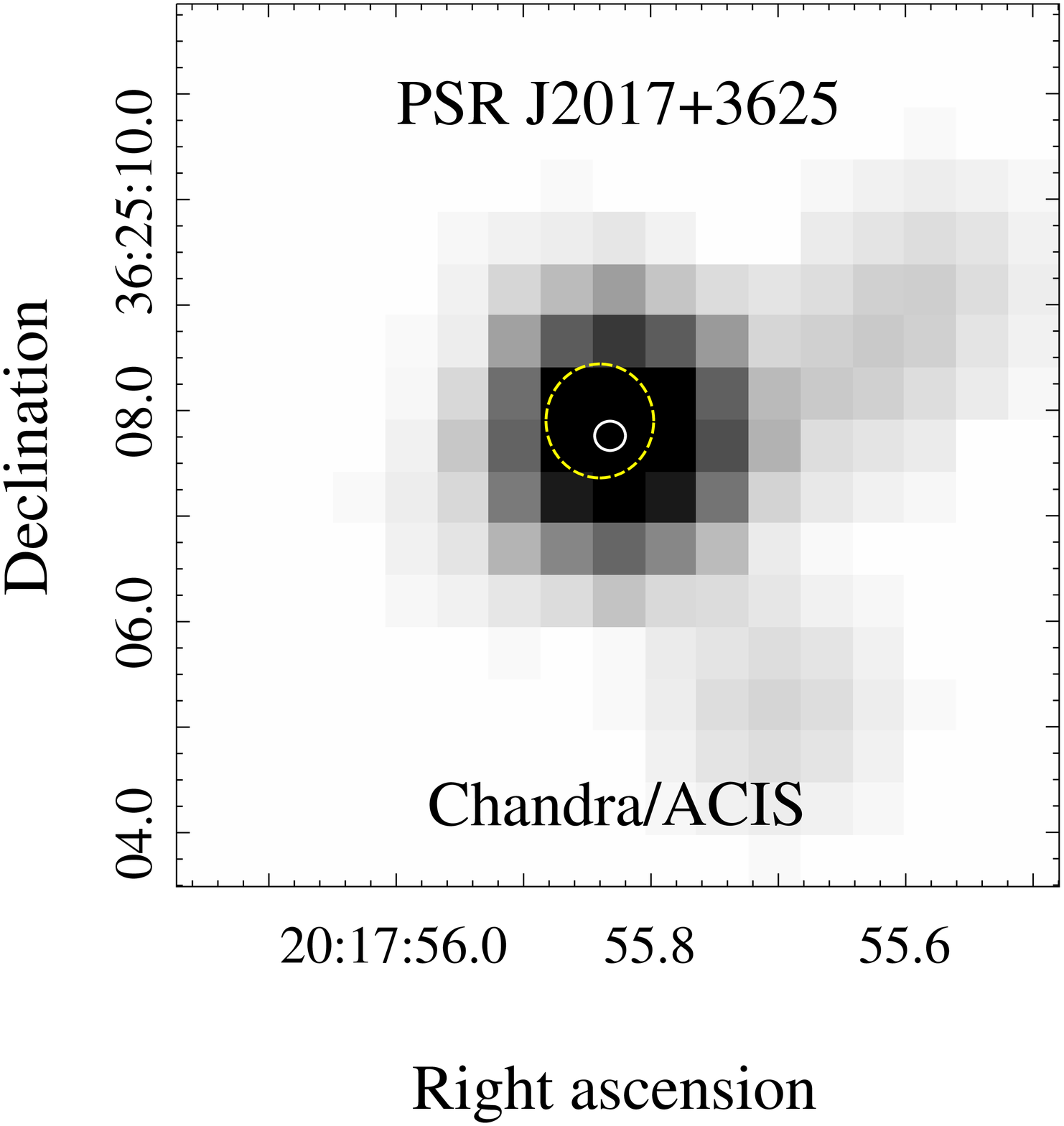}} 
   \end{picture}
  \end{center}
\caption{ \textit{Swift}/XRT and \textit{Chandra}/ACIS images of the point-like X-ray counterpart candidates for the $\gamma$-ray pulsars 
listed in Table \ref{t:psr-coo}. 1$\sigma$ positional uncertainties of the point-like candidate centroids and the $\gamma$-ray pulsars 
in the images are shown by solid white and dashed yellow  
ellipses, respectively. The green annulus in the bottom-left panel was used to extract the spectrum of the extended emission around 
the point-like counterpart candidate. The insert in this panel shows the zoomed-in 8 arcesec $\times$ 8 arcsec region with the candidate  
near  the centre of the extended emission. A different count scale is used in the insert to reveal the point-like counterpart candidate.} 
\label{f:swift-chandra-ima}
\end{figure*}

The sections of the stacked/merged  X-ray 
images containing the pulsars are presented in Fig.~\ref{f:swift-chandra-ima}  where dashed yellow and solid white  ellipses show 1$\sigma$ position uncertainties of 
the $\gamma$-ray pulsars and centroids of their X-ray counterpart candidates, respectively. The former account for 
the \textit{Fermi} 1$\sigma$ timing position uncertainties  from Table~\ref{t:psr-coo} and the X-ray telescope 1$\sigma$ astrometric accuracy. This quantity, in turn, 
is estimated from 
90\% confidence aspect solution accuracy
of 3\farcs5 and 0\farcs8 for  \textit{Swift} \citep{SwiftCatalogue}        
and \textit{Chandra}\footnote{cxc.harvard.edu/cal/ASPECT/almen}, respectively. The solid white ellipses include only  statistical point source position measurement  
errors on the stacked/merged images. It is seen that 
positions of the counterpart candidates are consistent with those of the pulsars in all cases.  

Due to shallow total exposures, mainly of about 10 ks, the formal  X-ray signal-to-noise ratio $S/N$ 
for  the counterpart candidates  
on the XRT  images is low. It  
varies from 1.8, for J0002$+$6216, to 5.9, for J1844$-$0346 (see the last column in Table~\ref{t:psr-coo}). Such a low $S/N$ 
is the main reason for relatively large absolute 
position errors as compared to 
the \textit{Fermi} position errors (cf. Table~\ref{t:psr-coo}).  Nevertheless, all the objects are classified as `Good' point sources 
in the \textit{Swift} 
catalogue based on an iterative point spread function and likelihood analysis \citep{SwiftCatalogue}.  $S/N$ 
for the ACIS counterpart candidates of J2017$+$3625 and  J0359$+$5414 is $\approx$5.5 and $\approx$23.8, respectively. 

As one can see in Fig.~\ref{f:swift-chandra-ima}, there is a diffuse emission  around the point-like candidate of  J0359$+$5414  extended roughly along 
SE--NW direction by about 30 arcsec. 
This can be the PWN, which is not a surprise  for the pulsar with a relatively 
high spin-down power of $\approx 10^{36}$ erg s$^{-1}$. 
The visible size of the presumed PWN ($\approx$ 30 arcsec $\times$ 20 arcsec) 
is roughly compatible with that of the Vela PWN ($\approx$ 6 arcmin $\times$ 5.5 arcmin) assuming 
that  the latter is moved   
to  about 12 times larger distance corresponding to J0359+5414 (from 0.3 to 3.5 kpc).

For further spectral analysis 
of the proposed counterpart candidates detected with \textit{Swift}, we have obtained their spectra using the \textit{Swift}/XRT 
data product generator \citep{evans2009}. For two \textit{Chandra} candidates, the spectra were extracted from the archival data using the 
{\sc ciao specextract} tool. For the J0359$+$5414 candidate, we have used the extraction aperture with a 2 arcsec radius for the point-like source,  
whereas for the extended source, we have used  
the annulus  centred at the point-like candidate with the inner and outer radii of 2 and 15 arcsec, respectively. 
The annulus is shown in Fig.~\ref{f:swift-chandra-ima}. 
For the J2017$+$3625 candidate,  a 1.5 arcsec aperture radius was used. In all the cases, the background was taken from  regions free of any sources.  

The  spectra were fitted 
in the 0.3--10 keV range
using the {\sc xspec} v.12.9.0 tool by plausible absorbed spectral models. We applied a power law (PL) 
model, describing  nonthermal radiation from the NS magnetosphere,  a blackbody (BB) model and a magnetised NS atmosphere models 
NSMAXG \citep{Ho2008}, describing the thermal emission from the NS surface. 
Depending on the count statistics, we used either a single model or a composite model.
The latter is the sum of the thermal and nonthermal spectral components, e.g.
BB+PL or NSMAXG+PL.
For the  diffusive source, we applied the  PL model assuming the synchrotron nature of the PWN emission.  
For the photoelectric absorption, the \texttt{tbabs} model with the \texttt{wilm} interstellar abundances was used. 
Due to low count numbers $C$-statistics \citep{1979cash} was used to estimate fit qualities.
Unless stated otherwise, all the spectra were binned by at least 1 count per energy bin.  
Independent information on the absorbing column density N$_H$ towards a pulsar was employed, when this was required to facilitate the  fit.  

\section{Properties of the pulsar X-ray counterpart candidates}
\label{prop}
\subsection{PSR J0002+6216}
\label{J0002}
\begin{figure}
\begin{center}
\includegraphics[width=9 cm, angle=0]{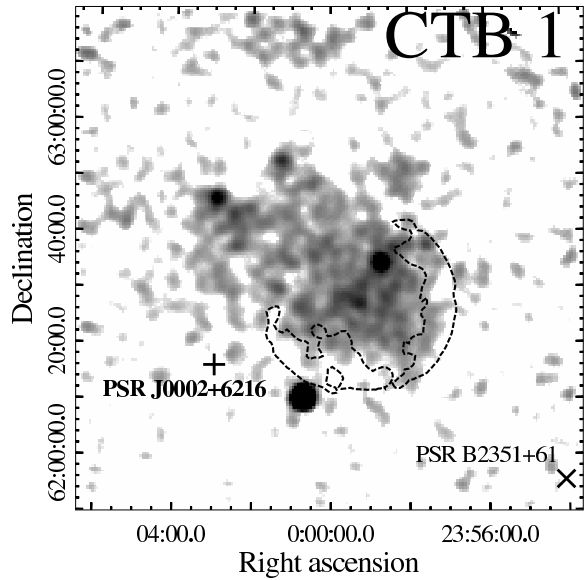}
\end{center}
\caption{\textit{ROSAT} X-ray image of the field containing the SNR CTB~1, 
PSR J0002$+$6216,  and  unrelated PSR B2351$+$61 whose positions 
are marked by a cross and an `X', respectively. 
Contours show the CTB 1 radio shell structure obtained from    
the Westerbork Northern Sky Survey \citep[WENSS;][]{WENSS}. 
}
\label{f:ctb1}
\end{figure}
The counterpart candidate for this pulsar accounts for 
only 9 XRT source counts detected during the 
$\approx$ 9.2 ks total exposure. 
This precludes any quantitative spectral analysis.

$N_{\rm H}$ can be estimated  if the pulsar is associated with the well studied  SNR CTB~1 (G116.9$+$0.2). 
The \textit{ROSAT} X-ray image of CTB~1 is presented in  Fig.~\ref{f:ctb1}. 
This is an oxygen-rich mixed morphology core-collapse remnant   
where a black-hole or an NS are expected to be born at the supernova event  
but none  have  been found so far \citep{Pannuti2010-ctb1,Pannuti2010-ctb1-xmm}. 
As seen from Fig.~\ref{f:ctb1}, PSR J0002$+$6216 is located near the edge of the remnant 
at the angular distance of only about 17 arcmin from its centre. 
Another nearby pulsar B2351$+$61 shown in the image cannot be associated with CTB~1 
because its trail calculated using  known proper motion 
and characteristic age \citep{harrison1993} does not point to the remnant. 
The PSR J0002$+$6216   pseudo-distance (Table~\ref{t:pulsars}) is consistent with the SNR
distance range of 1.5--4 kpc \citep[e.g.,][]{Fesen1997-ctb1} favouring the association.

The age of CTB~1 is estimated to be in the range of 7.5--44 kyr \citep{koo1991,hailey1994}. 
This is smaller than the pulsar characteristic age of 306 kyr.
However, such a large discrepancy is frequently observed for  pulsars associated with SNRs 
(e.g., PSR J0538+2817 in the SNR S147 \citep{Kramer2003}; 
PSR J1101$-$6101 in the SNR MSH 11$-$61A \citep{Halpern2014}). 
If the real pulsar age is consistent with that of the SNR, then  the initial pulsar period $P_0$  
according to the standard real age $t$--$P\dot{P}$ relation,   
$$t = {P \over (n-1)\dot P}\left[1-\left(P_0 \over P\right)^{n-1}\right],$$ 
was in the range of 107--114  ms for the magnetic dipole spin-down model with $n=3$.  
This is a typical value for birth periods \citep[see, e.g.,][]{popov2012,noutsos2013} which is also
in a good agreement with recent core-collapse models \citep[e.g.,][]{Wongwathanarat2013,fuller2015}.

Based on the 17 arcmin pulsar offset from the CTB~1 centre,  its expected 
proper motion and transverse velocity are $\mu\approx50t^{-1}_{\rm 20}$ mas~yr$^{-1}$ and 
$v_{\perp}\approx550d_{\rm 2.3}t^{-1}_{\rm 20}$ km~s$^{-1}$, where $d_{\rm 2.3}$ 
is the pulsar distance scaled to its pseudo-distance of 2.3 kpc (Table~\ref{t:pulsars})
and $t_{\rm 20}$ is the real age divided by  20 kyr.  
Given the   age uncertainty of 7.5--44 kyr, the $v_{\perp}$ uncertainty 
of (250--1500)$d_{\rm 2.3}$ km~s$^{-1}$  is large, 
but it is  compatible with the mean 2D pulsar speed of 
$\approx$ 250 km~s$^{-1}$ \citep{hobbs2005} and the largest NS velocity 
of $\approx$ 1100 km~s$^{-1}$  precisely measured so far \citep{chatterjee}. 

Therefore, the association of  PSR J0002$+$6216 and CTB~1 appears to be plausible. 
Assuming that it is real, we fixed $N_{\rm H}$ for the pulsar at $6 \times 10^{21}$~cm$^{-2}$, 
the value accepted for the remnant \citep{Pannuti2010-ctb1}. Then, 
to estimate the pulsar counterpart candidate flux, we  
fitted its spectrum with the absorbed PL with the photon index $\Gamma=2$, 
which is typical for pulsars \citep[e.g.,][]{kargaltsev2008}. The resulting $C$-value was 12 per 6 degrees of freedom (d.o.f.). 
The fit resulted in the observed and unabsorbed X-ray fluxes
of $4.4^{+2.8}_{-2.0} \times 10^{-14}$   
and $7.4^{+4.6}_{-3.9} \times 10^{-14}$ erg~cm$^{-2}$~s$^{-1}$,
respectively\footnote{Hereafter, 1$\sigma$  flux and fitting parameter errors are presented.}, in the 0.3--10 keV band.
For the distance of 2.3 kpc this translates into the X-ray luminosity of 
$4.7^{+2.9}_{-2.5} \times 10^{31}d_{\rm 2.3}^2$ erg~s$^{-1}$. 
It is consistent with non-thermal X-ray luminosities of 
10--100 kyr old pulsars \citep[e.g.,][]{kargaltsev2008}.
The X-ray efficiency in the  0.3--10 keV range is Log$(L_{\rm X}/\dot{E})\approx -3.5$.
The ratio between the $\gamma$-ray and unabsorbed X-ray fluxes is 
Log$(G_{100}/F_{\rm X})\approx 2.4$. According to 
\citet[][see their Fig.~18]{abdo2013}, this is consistent with a typical  for  radio-loud 
$\gamma$-ray pulsars ratio of 2.4$\pm$1.1, encouraging  further search for the pulsar radio 
counterpart\footnote{After this paper has been submitted, the detection of J0002$+$6216 
in the radio was reported \citep{wu2017}.}. 

Alternatively, the single BB fit results in  the temperature $T=147^{+84}_{-43}$ eV,  the emitting area radius $R=1.3^{+3.3}_{-1.2} d_{2.3}$ km, and 
the  observed and unabsorbed X-ray fluxes  of $1.5^{+0.8}_{-0.6} \times 10^{-14}$ and $1.2^{+2.4}_{-0.8} \times 10^{-13}$ erg~cm$^{-2}$~s$^{-1}$, respectively. 
The fit quality was only marginally better than that for the PL model, $C({\rm d.o.f})=9.6(5)$. In principle, this result could be consistent with the thermal emission from pulsar hot polar caps. However, uncertainties are too large for definite conclusions.      

To summarise, we can state that 
the detected X-ray source is a plausible counterpart of PSR J0002$+$6216.

\subsection{PSR J0359+5414}
\label{J0359}
 
To reveal the nature of the serendipitously  detected  \textit{Chandra} pulsar and PWN  counterpart candidates,   
we have used the {\sc combine\_spectra} tool to co-add the source and background spectra from all the observation IDs. 
This resulted  in 150 and 293 ACIS source counts for the point and the diffuse objects, respectively, detected  during the $\approx$460 ks total 
exposure in the 0.3--10 keV range.
The spectra were grouped to ensure 5 and 10 counts per the energy bin 
for the presumed pulsar and  PWN, respectively.  
We fitted the  spectra  simultaneously with the  shared $N_{\rm H}$ parameter using 
all   spectral models listed in Sect. \ref{search}.

\begin{figure}
  \setlength{\unitlength}{1mm}
  \begin{center}
    \includegraphics[scale=0.37, angle=0]{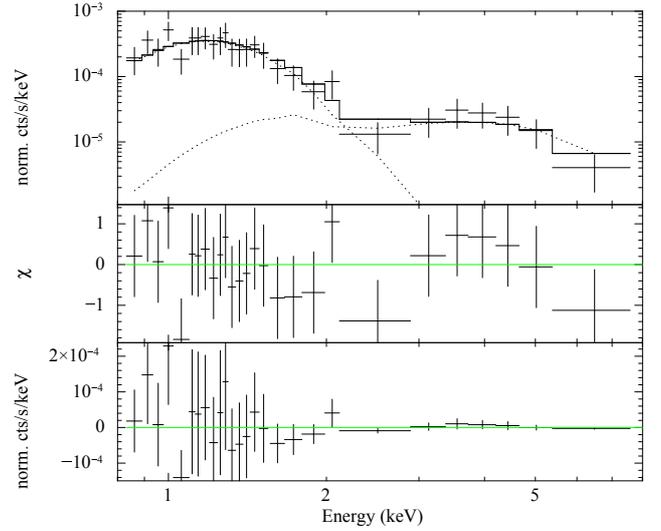}
  \end{center}
  \caption{{\sl Top panel}: X-ray spectrum of the PSR J0359$+$5414 X-ray counterpart candidate 
  best fitted by the sum (solid line) of the NSMAXG and the PL models which are dominated in 
  the 0.5--2 and 2--8 keV ranges, respectively (dotted lines).
  {\sl Middle and bottom panels}: fit residuals.}
  \label{fig:spectrum-psrj0359}
\end{figure}

\begin{figure}
  \setlength{\unitlength}{1mm}
  \begin{center}
      \includegraphics[scale=0.37]{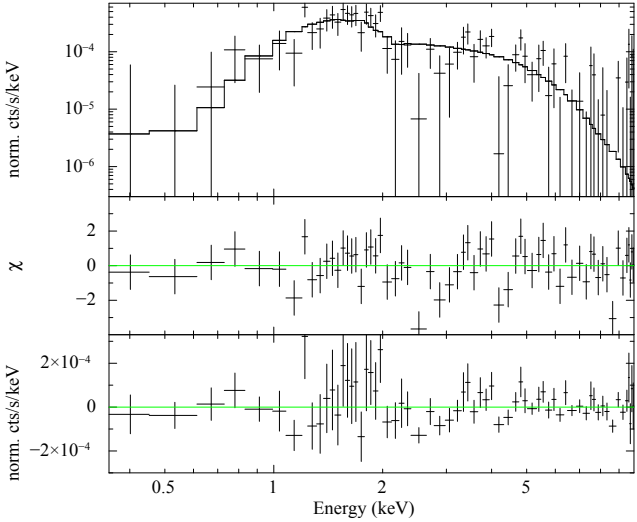}
  \end{center}
  \caption{{\sl Top panel}: X-ray spectrum of the presumed PWN of PSR J0359$+$5414 and 
  the best-fit PL model (solid line).
  {\sl Middle and bottom panels}: fit residuals.}
  \label{fig:spectrum-psrj0359pwn}
\end{figure}

For the pulsar counterpart candidate only the composite models BB$+$PL and NSMAXG$+$PL were statistically acceptable.  
Both of them provided equal fit qualities, $C_{\rm BB+PL}({\rm d.o.f.})=89.8(85)$ and $C_{\rm NSMAXG+PL}({\rm d.o.f.})=89.95(86)$.  
For the PL$+$NSMAXG model (with NS mass $M_{\rm NS}=1.4{\rm M_\odot}$, radius $R_{\rm NS}=13$ km and 
magnetic field $B=10^{12}$ G), fixing the distance at the pseudo-distance value of 3.45 kpc, 
we have obtained the column density $N_{\rm H}=(0.9\pm0.2)\times 10^{22}$~cm$^{-2}$,
the photon index $\Gamma=0.5\pm 0.6$ and 
the NS temperature (gravitationally redshifted) $T^{\infty}=(6.1\pm 0.2)\times 10^5$ K ($\approx53$ eV).
For the BB$+$PL model, the column density $N_{\rm H}=(0.9\pm 0.3) \times 10^{22}$~cm$^{-2}$,
the photon index $\Gamma=0.8\pm0.6$, the temperature $T=162^{+29}_{-25}$ eV and 
the emitting area radius $R=0.8^{+1.1}_{-0.4}d_{\rm 3.45}$ km, 
where $d_{\rm 3.45}$ is the pulsar distance normalised to  the pseudo-distance. 
In this case, the thermal emission comes from 
a compact hot region on the NS 
surface whose radius is consistent with the polar cap radius of 0.76 km
estimated  for this pulsar in the standard way \citep[e.g.,][]{sturrock1971}.
For the presumed PWN, in both cases we got $\Gamma=1.6\pm0.3$, 
which is typical for PWNe \citep{kargaltsev2008}. 
For the NSMAXG$+$PL case, the point-like and the extended object spectra and the respective best fit  models 
are presented in Figs.~\ref{fig:spectrum-psrj0359} and \ref{fig:spectrum-psrj0359pwn}. 

Our spectral analysis strongly supports the pulsar nature of the counterpart candidate. 
Its X-ray spectrum is consistent with the spectrum expected from this middle-aged 
pulsar with the soft thermal component from the entire surface of the cooling NS, 
or from its hot regions, and a hard nonthermal spectral tail of the pulsar magnetosphere origin. 
 
The absorbing column density value for PSR J0359$+$5414 
derived from the spectral fit is consistent with $N_{\rm H}\approx 1.1\times10^{22}$~cm$^{-2}$ 
estimated towards the pulsar for the pseudo-distance of 3.45 kpc 
using the relation $N_{\rm H} = (8.9\pm0.4)\times E(B-V)\times 10^{21}$~cm$^{-2}$ 
\citep{foight2016} and the interstellar extinction $E(B-V)$--distance relation  
which is based on the Pan-STARRS and 2MASS photometry\footnote{http://argonaut.skymaps.info/}
 \citep{dustmap2015}. 
This means that 3.45 kpc is a quite realistic distance estimate.  
 
For the pulsar and PWN spectra, the unabsorbed nonthermal fluxes  
in the 0.3--10 keV range are  $(9\pm3)\times10^{-15}$  and 
$(2\pm0.3)\times10^{-14}$ erg~cm$^{-2}$~s$^{-1}$, respectively.
These translate into nonthermal luminosities $L_{\rm X}^{\rm psr}\approx1.3\times10^{31} d^2_{\rm 3.45}$ 
and $L_{\rm X}^{\rm pwn}\approx2.8\times10^{31} d^2_{\rm 3.45}$ erg~s$^{-1}$ which are compatible with 
known PSR/PWN luminosities of pulsars with $\dot{E}\sim10^{36}$ erg~s$^{-1}$ \citep{kargaltsev2008}.
Finally, the ratio between the $\gamma$-ray and the nonthermal unabsorbed X-ray fluxes of the pulsar is 
Log$(G_{100}/F_{\rm X})\approx 3.4$. This is consistent with the value of $3.5\pm0.5$ 
derived for the radio-quiet population of $\gamma$-ray pulsars \citep{abdo2013}. 

To summarise, the X-ray properties of the \textit{Chandra}  point and extended sources leave little doubts that they represent   
the real X-ray counterparts of PSR J0359$+$5414 and its PWN.

\subsection{PSR J0554+3107}
\label{J0554}
  
There are 17 source counts from the XRT counterpart candidate detected during the 
$\approx$10 ks total exposure in the 0.3--10 keV range.    
This precludes any 
 quantitative spectral analysis
using the XRT data alone. 
Nevertheless,  since 13 out of  17 counts are detected in the 0.3--1 keV band,  a preliminary conclusion can be drown  
that the source spectrum is soft. 
This suggests that the detected emission may be  dominated 
by the thermal component from the NS surface, as is typically observed for middle-aged pulsars, a class  
to which PSR J0554+3107 likely belongs in  view of   its characteristic age (Table~\ref{t:pulsars}).

This is further  supported by the pulsar association with 
an evolved SNR G179.0+2.6 proposed by \citet{Pletsch2013}. 
The pulsar characteristic age of 52 kyr is compatible 
with the remnant age range of 10--100 kyr \citep{Fuerst-Reich1986}.  
Detection of G179.0+2.6 in the optical O\thinspace{\sc iii} narrow 
band\footnote{http://astro.neutral.org/imagehtml/20161128-snr-G179.0+2.6.html}
implies that this is an oxygen rich  core-collapse SNR where a NS could have been  born. 
Based on an empirical surface luminosity--SNR angular size relation \citep{milne1979} 
and $d_{\gamma}$ for the pulsar (Table~\ref{t:pulsars}), the likely distance 
to the presumed pulsar+SNR system lies in the range of 1.9--3.5 kpc.  

This distance range, the $E(B-V)$--distance \citep{dustmap2015} and 
$E(B-V)$--$N_{\rm H}$ \citep{foight2016} relations allow us to constrain $N_{\rm H}$ for the pulsar 
to the range of $(1.9-2.5)\times 10^{21}$ cm$^{-2}$ and perform 
a rough spectral analysis of the counterpart candidate. 
Fixing $N_{\rm H}$ in the obtained range 
we have fitted the XRT source spectrum in the 0.3--10 keV range 
by the PL, BB, and  NSMAX  models. For the latter model, we have selected      
 $B=10^{13}$~G and the gravitational redshift parameter $1+z_g=1.21$, 
corresponding to $M_{\rm NS}=1.4{\rm M_\odot}$ and $R_{\rm NS}=13$ km.  

The PL fit results in an unreasonably steep spectral slope with  $\Gamma>4$. This is not typical 
for pulsars and thus can be rejected. Both thermal fits are equally acceptable with $C({\rm d.o.f.})=15(12)$. 
The BB fit results in the temperature of 100--110 eV and the emitting area radius of (2--3)$d_{\rm 1.9}$ km
depending on the $N_{\rm H}$ value from the allowed range. 
Here $d_{\rm 1.9}$ is the pulsar distance in units of 1.9 kpc. 
The NSMAX fit yields  the NS surface temperature and 
emitting area radius (gravitationally redshifted) of 47--54 eV and (11--19)$d_{\rm 1.9}$ km, respectively.  
For both thermal models the absorbed flux is about $(3.1-3.2)\times10^{-14}$ erg~cm$^{2}$~s$^{-1}$
in the 0.3--10 keV band. 
 
The  temperature and the emitting area derived  from the fits using the thermal emission models  
are consistent with those expected for a middle-aged cooling NS. 
This supports the pulsar nature of the XRT  counterpart candidate.

\subsection{PSR J1105$-$6037}
\label{J1105}
There are 16 source counts from the XRT counterpart candidate detected 
 during the $\approx$16 ks total exposure in the 0.3--10 keV range. 
10 out of 16 counts are  in 0.3--1 keV range 
 implying, as in the  case of PSR J0554$+$3107, 
that the spectrum may contain a soft thermal component 
corresponding to the surface emission from  a middle-aged NS. 
 
Unfortunately, no advanced interstellar extinction--distance relation
is provided for the pulsar direction by \citet{dustmap2015} to constrain 
$N_{\rm H}$ to the pulsar and to facilitate spectral fits. 
Therefore we have utilised the extinction map 
which is based on a model fit to the COBE DIRBE data \citep{Drimmel2003}.
Applying the $E(B-V)$--$N_{\rm H}$ relation \citep{foight2016} 
we obtained $N_{\rm H}\approx 3.2\times 10^{21}$~cm$^{-2}$ 
for the pulsar pseudo-distance of 1.53 kpc (Table~\ref{t:pulsars}).

Fixing $N_{\rm H}$ at this value,  we have fitted the source spectrum
trying the PL, BB and NSMAXG models.
The PL fit resulted in a too steep photon index $\Gamma\approx4$ 
which is not typical for pulsars. The nonthermal model  can thus be rejected. 
Both thermal models resulted in acceptable fits with $C_{\rm BB}({\rm d.o.f.})=5.7(9)$ and $C_{\rm NSMAXG}({\rm d.o.f.})=6.3(10)$.
Using the BB model, we obtain the temperature of $165^{+58}_{-38}$ eV and 
the radius of the emitting area of $0.5^{+0.6}_{-0.3}d_{1.53}$ km, where $d_{1.53}$ is
the pulsar distance in units of the pseudo-distance. 
The latter value is consistent with the polar cap radius of 0.49 km estimated for this pulsar in the standard way \citep[e.g.,][]{sturrock1971}.
For the NSMAXG model (the NS mass $M_{\rm NS}=1.4{\rm M_\odot}$, radius $R_{\rm NS}=13$ km and 
magnetic field $B= 10^{12}$ G), fixing the distance at the pseudo-distance value,
we obtain the gravitationally redshifted temperature of $(5.1\pm0.3)\times10^{5}$~K ($\approx 44$ eV).
The observed flux in the 0.3--10 keV band is $\approx 2\times10^{-14}$ erg~cm$^{2}$~s$^{-1}$ for both thermal models. 

The thermal fit parameters are reasonable for a middle-aged cooling NS
supporting the pulsar nature of the XRT candidate counterpart.

\begin{figure}
\setlength{\unitlength}{1mm}
\begin{center}
\includegraphics[scale=0.36, angle=0]{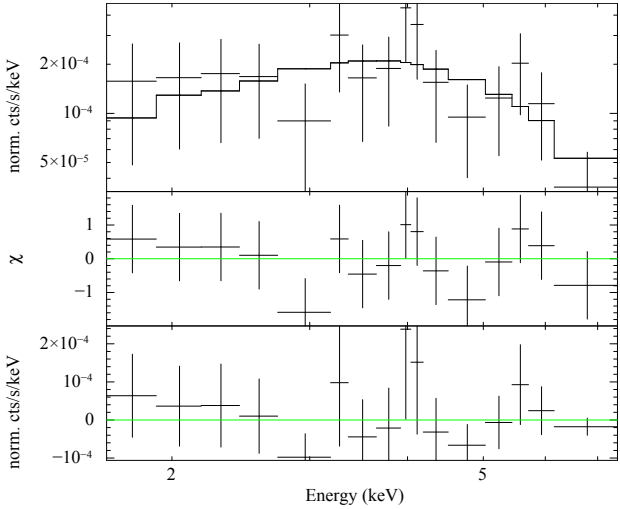}
\end{center}
\caption{{\sl Top panel}: X-ray spectrum of the PSR J1844$-$0346 X-ray counterpart candidate
and the best-fit PL model. The spectrum is binned to ensure at least 4 counts per energy bin
for illustration purposes only. {\sl Middle and bottom panels}: residuals}
\label{fig:spectrum-psrj1844}
\end{figure}
\subsection{PSR J1844$-$0346}
\label{J1844}

This  is the youngest and most energetic  
pulsar in our sample but it is likely to be the most distant object as well (Table~\ref{t:pulsars}). 
About 100 source counts 
 during the
 $\approx$100 ks total exposure were detected from the XRT candidate allowing one to 
perform an initial spectral analysis without additional assumptions. 
The spectrum is well fitted by the  
PL model with $N_{\rm H}=5^{+4}_{-3}\times 10^{22}$ cm$^{-2}$
and $\Gamma=0.9^{+0.9}_{-0.8}$; $C_{\rm PL}({\rm d.o.f.})=44.2(61)$. 
This suggests  that the candidate can be 
indeed a distant young pulsar whose spectrum is dominated by 
the nonthermal emission of the NS magnetosphere origin, as is observed, e.g., for the Crab pulsar.
The BB fit is also acceptable but results in an unrealistically high for NSs temperature $\ga$ 1.5 keV.  
Due to this reason, the thermal model can be rejected.    
The spectrum with the PL fit  is  shown in Fig.~\ref{fig:spectrum-psrj1844}.  
The observed and unabsorbed fluxes are 1.6$^{+0.5}_{-0.4}\times 10^{-13}$  
and $2.2^{+1.3}_{-0.4}\times 10^{-13}$ erg~cm$^{-2}$~s$^{-1}$, respectively,  in the 0.3--10 keV band.
This yields the X-ray luminosity of 4.9$^{+2.9}_{-0.9}\times 10^{32} d^2_{4.3}$ erg~s$^{-1}$, 
where $d_{4.3}$ is the pulsar distance in  units of the pseudo-distance.
Log$(G_{100}/F_{\rm X})\approx 2.1$ 
is compatible with the value 
obtained for the radio-loud $\gamma$-ray pulsar population \citep{abdo2013}. 
\begin{figure}
\begin{center}
\includegraphics[scale=0.27, angle=0,clip]{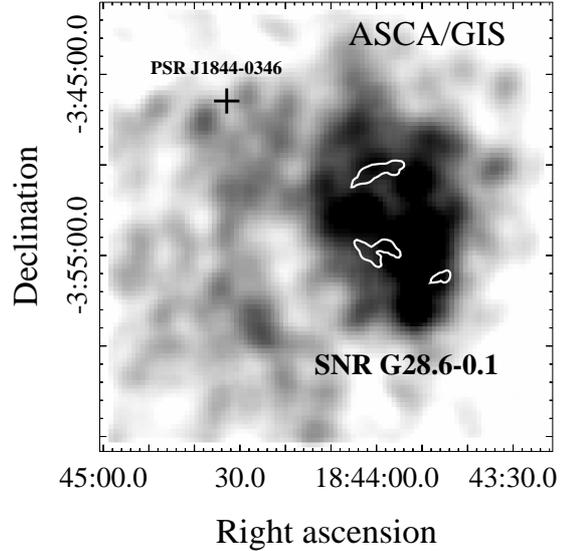}
\end{center}
\caption{\textit{ASCA}/GIS X-ray image of the field containing the SNR G28.6$-$0.1 and PSR J1844$-$0346
whose position is marked by a cross. 
Contours show the G28.6$-$0.1 radio structure obtained from    
the MAGPIS (The Multi-Array Galactic Plane Imaging Survey) 
VLA 1.4 GHz data \citep{helfand2006}. }
\label{f:j1844_snr}
\end{figure} 

 PSR J1844$-$0346 is located about 10 arcmin  off the  G28.6$-$0.1 SNR centre (Fig.~\ref{f:j1844_snr}). 
This is a young 
SNR thoroughly explored by \citet{Ueno2003} with \textit{Chandra}. 
They obtained the column density of $(3.2-4.5)\times 10^{22}$ cm$^{-2}$ to the remnant, which is compatible with what 
we obtained for the pulsar counterpart candidate.
The SNR X-ray radiation is dominated by synchrotron emission from an expanding shell.  
It exhibits a clumpy morphology with an overall elliptical shape with 
the NE major axis orientation. 
The radio and X-ray emission regions are not correlated \citep{Bamba2001}, which is typical for mixed morphology SNRs.  
The distance to the remnant and its age were recently estimated to be  $9.6\pm0.3$ kpc  and 19 kyr, respectively \citep{2017Ranasinghe}.

No pulsar detection has been reported within the remnant, while  PSR J1844$-$0346 is located near 
its NE boundary which has not been observed by \textit{Chandra}. 
The  association with the SNR implies the pulsar proper motion $\mu=50t^{-1}_{12}$ mas~yr$^{-1}$
and the transverse velocity $v_{\perp}\approx1020d_{\rm 4.3}t^{-1}_{\rm 12}$ km~s$^{-1}$, 
where $t_{\rm 12}$ is the pulsar age scaled by 
its characteristic age of 12 kyr (Table~\ref{t:pulsars}).
This velocity is close to the
highest firmly established NS velocity of $\approx$ 1100 km~s$^{-1}$ \citep{chatterjee}.
If the real age and distance of the presumed PSR+SNR system  are equal to those estimated for the remnant,  
the transverse velocity becomes higher, $1440\pm 50$ km~s$^{-1}$, but it is still not outstanding.  
 
\subsection{PSR J2017$+$3625}
\label{J2017}
This is the oldest and least energetic but likely closest pulsar in our sample (Table~\ref{t:pulsars}).
\citet{gotthelf2016} failed to identify    
it in X-rays in the same \textit{Chandra} data-set
using  the pulsar coordinates from the 3FGL catalogue, 
which are different from more accurate coordinates obtained from the timing analysis
by \citet{Clark2017}. There are 10 ACIS source counts from the  counterpart candidate detected during the 
$\approx$10 ks exposure in the 0.3--10 keV range.

We have estimated $N_{\rm H}$ to the pulsar using the interstellar
extinction--distance relation \citep{dustmap2015} and the pulsar 
pseudo-distance of 0.656 kpc (Table~\ref{t:pulsars}).  
The resulting $N_{\rm H}$ is about $5.3\times10^{20}$~cm$^{-2}$ which was then frozen in spectral fits.   
The spectrum was fitted by the  PL and BB models.

For the BB model we have obtained  the temperature of about 1 keV which is extremely high for a thermally emitting isolated NS,  particularly 
for this 2 Myr old pulsar. We have thus rejected this model. 
For the PL model the fit is acceptable, $C({\rm d.o.f.})=7.7(8)$, and it results in  $\Gamma=1.2\pm0.6$.  
This is  consistent with the spectral slopes of the nonthermal X-ray emission detected in other pulsars. 
The observed and unabsorbed fluxes in the 0.3--10 keV energy range 
were $2.2^{+1.4}_{-0.8}\times 10^{-14}$ 
and  $2.3^{+1.3}_{-0.8}\times 10^{-14}$ erg~cm$^{-2}$~s$^{-1}$, respectively. 
The latter translates in the X-ray luminosity 
$L_{\rm X}=1.2^{+0.7}_{-0.4}\times 10^{30}d^2_{0.656}$ erg s$^{-1}$.
The corresponding X-ray efficiency Log($L_{\rm X}/ \dot{E})\approx -4$. 
The ratio Log$(G_{100}/F_{\rm X})\approx 3.45$ 
is consistent with the value of $3.5\pm0.5$ 
derived for the radio-quiet $\gamma$-ray pulsars \citep{abdo2013}.  

The detected  \textit{Chandra} source appears to be a plausible X-ray counterpart 
of this rather old pulsar.


\section{Possible optical counterparts to the X-ray sources}
\label{oc}
The X-ray count statistic is too low  
for five of the six counterpart candidates
 to confirm the 
positional association with gamma-ray pulsars through their spectral properties.
Therefore, it is a good idea to check if they  coincide positionally 
with unrelated 
background  optical objects which could emit in X-rays and thus lead to a false pulsar identification. 
To do that, we have  used the \textit{Swift} Ultraviolet/Optical Telescope (UVOT) images and the results of  deep, up to a $\approx$23 visual magnitude 
limit,  optical sky surveys 
including the pulsar fields. Specifically, we have considered      
the recent optical databases of the Panoramic Survey Telescope and Rapid Response System 
\citep[Pan-STARRS;][]{panstarrs1}\footnote{https://panstarrs.stsci.edu/}, 
and VST Photometric H$\alpha$ Survey of the Southern Galactic Plane and Bulge 
\citep[VPHAS+;][]{vphas}\footnote{http://www.eso.org/sci/observing/PublicSurveys.html}. 

We have not found any optical objects down to $\approx$23.3 visual magnitude limit within the position error ellipse  
of the  \textit{Chandra} counterpart candidate for J2017$+$3625 (Table~\ref{t:psr-coo}). This favours the correct X-ray identification of the pulsar.

Within the position error ellipse of the  \textit{Swift} counterpart candidate for J1844$-$0346  with purely nonthermal X-ray spectrum ({Sect.~\ref{J1844}),  
we have found a relatively bright optical source PSO J184433.074$-$034628.369 with magnitudes $g=19.729$, $r=18.169$, $i=17.243$, $z=16.715$.
By its colour, it appears to be an M-class main sequence star which can hardly produce such a hard X-ray spectrum as is observed for the candidate. 
We have  also not found any  known active galactic nuclei at our candidate position which could be responsible  
for the nonthermal X-ray spectrum.

We  have found  a relatively bright background star VPHAS J110500.49$-$603713.41 with $u'=17.96$, $g'=16.97$, $r'=16.33$, $i'=16.1$  
within the position error ellipse of the J1105$-$6037 \textit{Swift}  X-ray counterpart candidate. 
Its spectral energy distribution is typical for  main-sequence stars. 

There are two faint red point objects PSO J055404.767+310739.519 with $i=21.655$  
and PSO J055405.198+310742.124 with $i=21.918$  within the position ellipse  of  
the  \textit{Swift}  counterpart candidate for J0554$+$3107. They are not detected in other optical bands. In the absence of any information 
on the object colours,  their nature  remains unknown.  

Finally, within the position error ellipse of the  \textit{Swift} counterpart candidate for J0002$+$6216  there are two 
relatively bright background point optical sources  PSO J000257.781+621607.150 with  $g=18.9$, $r=17.9$, $i=17.42$, $z=17.17$ and $y=17$, 
and PSO J000257.830+621613.565 with  $g=17.76$, $r=16.9$, $i=16.49$, $z=16.29$ and $y=16.1$. Both sources are likely  main sequence stars.

Therefore,  the X-ray identification of the latter three pulsars has to be viewed  with   caution. 
More sensitive, high spatial and temporal resolution X-ray observations can help  
to distinguish  possible false identifications from the real ones.

\section{Discussion and conclusions}
\label{dc}
We have found possible X-ray counterparts for six recently discovered 
$\gamma$-ray pulsars and have proposed SNR associations for two of them. 
All the counterpart candidates   coincide  by position well with the pulsars. 
They are located in the Galactic plane and their X-ray fluxes are $\ge2\times 10^{-14}$ erg cm$^{-2}$ s$^{-1}$. 
According to the LogN--LogS distribution for the Chandra Galactic Plane sources from \citet{2005ebisava},  
the probability of chance detection of an unrelated X-ray source 
with $F_{\rm X}>2\times10^{-14}$ erg cm$^{-2}$ s$^{-1}$  at the 
$\gamma$-ray pulsar position is  $\lesssim 10^{-4}$. This small value supports the correct  X-ray identifications. 

Despite a low count statistics for most candidates, qualitative  conclusions  on X-ray spectral properties   
can be estimated for all of them. These properties are consistent with those observed for other 
rotation powered pulsars, which demonstrate either a non-thermal PL spectra of the NS magnetosphere origin or a thermal spectra  
from the NS surface, or a combination of the two.  
The estimated spectral indices and temperatures   are in a good agreement  with those observed 
for other pulsars with similar parameters. Moreover, the ratios of the $\gamma$-ray fluxes to the non-thermal X-ray fluxes, in cases where 
it is possible to estimate the latter, 
are also  compatible with the respective ratios for other $\gamma$-ray pulsars firmly identified in X-rays. 
Thus, the inferred X-ray properties  also favour the correct identification.

The most secure X-ray identification is  for PSR J0359$+$5414 ({Sect.~\ref{J0359}).  
Almost 500 ks of  \textit{Chandra} observations allowed us  
to reveal also the  diffuse 
emission around the pulsar. 
In view of its extent and  non-thermal spectrum, it is consistent   
with  PWN emission which can be  powered by this quite energetic pulsar. The pulsar counterpart spectrum shows  
a thermal component from the NS surface,  dominating at low photon energies, and a high-energy nonthermal tail of the NS magnetosphere origin, 
as is observed for Vela-like and middle-aged pulsars. Overall, there is little doubt that the \textit{Chandra} source CXOGSC J035926.0+541455 
is the real pulsar counterpart. 
The absorbing column density toward the pulsar 
derived from the X-ray spectral fits together with the available extinction--distance relation allowed us to estimate the distance 
to the pulsar as about 3.45 kpc. 
It is consistent with the $\gamma$-ray pseudo-distance value. 
The object is a very faint X-ray source.  
With current X-ray instruments only a mega-second dedicated exposure  can help to better constrain its spectral 
parameters and to find pulsations with the pulsar period.

Medium exposure observations of the other five pulsar counterpart candidates with \textit{Chandra} and \textit{XMM-Newton} 
are necessary  to  confirm firmly their NS nature. Improved count statistics  and  
detection of pulsations 
will help to exclude  their chance coincidence with 
background optical stars possibly emitting in X-rays.
The suggested associations of PSRs J0002+6216 and J1844$-$0346  with the SNRs CTB 1 and G28.6$-$0.1, respectively, can be confirmed by 
measuring the pulsar's proper motion using a set of \textit{Chandra} observations separated by a few years.  
Detection of bow-shock PWNe with tails pointing  to the centres of the
respective remnants  would also favour the associations.  In this case, the absorbing column densities to the pulsars derived from the advanced data 
have to be consistent with 
those of the SNRs.    

 After this paper submission, the work by \citet{wu2017}  appeared where authors using the same archival X-ray data briefly reported 
on five of the six X-ray  counterpart candidates thoroughly considered here.   However, they  presented only a simplified spectral analysis 
using a single PL model for all the candidates and used only about 10 per cent of the \textit{Chandra} data available 
for PSR J0359$+$5414.  The counterpart candidate for PSR J0554$+$3107 and possible pulsar associations with SNRs are not considered in that paper. 
The authors have not reported the  background optical objects within the X-ray position error ellipses, possibly emitting in X-rays,  
which may call into question
the correctness of the identifications, and have not estimated the probability of chance detection of  unrelated X-ray sources at the pulsar's positions.
At the same time, they have reported the detection of  PSR J0002+6216 in the radio, which is predicted by our X-ray analysis (Sect.~\ref{J0002}).   

\section*{Acknowledgements}

We are grateful to the anonymous referee for useful suggestions improving the paper and to D.A. Baiko 
for careful reading the manuscript and valuable comments. 
The work was supported by the Russian Science Foundation, grant 14-12-00316. 
It made use of data supplied by the UK Swift Science Data Centre at the University of Leicester, and data and/or 
software provided by the High Energy Astrophysics 
Science Archive Research Center (HEASARC), which is a service of the Astrophysics Science 
Division at NASA/GSFC and the High Energy Astrophysics Division of the Smithsonian Astrophysical Observatory.   
Based on observations made with ESO Telescopes at the La Silla or Paranal Observatories under programme ID(s) 
177.D-3023(B), 177.D-3023(C), 177.D-3023(D), 177.D-3023(E). 



\bibliographystyle{mnras}

\bibliography{fermi} 








\bsp	
\label{lastpage}
\end{document}